\RequirePackage{ifpdf}
\documentclass[12pt,letterpaper]{JHEP3}         % For preprint
\usepackage{amssymb,amsfonts}
\usepackage{epsfig}
\def\be{\begin{eqnarray}}
\def\ee{\end{eqnarray}}
\newcommand{\nn}{\nonumber}
\newcommand\para{\paragraph{}}

\newcommand{\eqn}[1]{(\ref{#1})}

\def\Dslash{\,\,{\raise.15ex\hbox{/}\mkern-12mu D}}
\def\Dbarslash{\,\,{\raise.15ex\hbox{/}\mkern-12mu {\bar D}}}
\def\delslash{\,\,{\raise.15ex\hbox{/}\mkern-9mu \partial}}
\def\delbarslash{\,\,{\raise.15ex\hbox{/}\mkern-9mu {\bar\partial}}}
\def\pslash{\,\,{\raise.15ex\hbox{/}\mkern-9mu p}}
\def\calDslash{\,\,{\raise.15ex\hbox{/}\mkern-12mu {\cal D}}}

\def\lae{\mathrel{\mathop{\smash{\lower .5 ex \hbox{$\stackrel<\sim$}}}}}
\def\lae{\mathrel{\mathop{\smash{\lower .5 ex \hbox{$\stackrel>\sim$}}}}}

\title{ADHM Revisited: Instantons and Wilson Lines}

\author{David Tong and Kenny Wong\\
Department of Applied Mathematics and Theoretical Physics, \\
University of Cambridge, \\
Cambridge, CB3 0WA, UK\\{\tt d.tong, k.wong@damtp.cam.ac.uk}}

\abstract{We revisit the well-studied D0-D4 system of D-branes and its relationship to the ADHM construction. It is well known that the D0-branes appear as instantons in the D4-brane worldvolume. We  add a Wilson line to the D4-brane in the guise of an extended fundamental string and determine how this affects the D0-brane dynamics. As the D0-brane moves in the presence of the Wilson line, it experiences a Lorentz force, proportional to its Yang-Mills gauge connection. From the perspective of the D0-brane quantum mechanics, this force emerges through the ADHM construction of the self-dual gauge connection.}

\begin{document}
\pagestyle{plain} \setcounter{page}{1}
\newcounter{bean}
\baselineskip16pt \setcounter{section}{0}

\section{Introduction}

The ADHM construction provides a beautifully simple method for finding solutions to the self-dual Yang-Mills equations, reducing non-linear partial differential equations to some straightforward linear algebra \cite{adhm}. 

%In 1978, Atiyah, Drinfeld, Hitchin and Manin described a beautiful and simple method for constructing instantons \cite{adhm}. Now known as the ADHM construction, it  reduces the problem of solving the self-dual Yang-Mills equations to some basic linear algebra.

\para
There is a long history of viewing the ADHM construction through the lens of string theory, starting with the work of Witten \cite{wittenadhm}. This provides a physically intuitive picture for how the construction works. The simplest approach involves D-branes in Type II string theory and the observation that D0-branes appear as instantons when nestled inside D4-branes \cite{douglas1}. The dynamics of the D0-branes is governed by a quantum mechanical gauge theory whose low-energy degrees of freedom can be thought of as the ADHM data. The space of ground states of this quantum mechanics coincides with the wavefunctions on the moduli space of instantons.

\para
While the D0-D4 quantum mechanics naturally describes  the instanton moduli space in terms of ADHM data, it does not tell us how to build the Yang-Mills gauge field itself. In other words, it doesn't capture the constructive part of the ADHM construction. For this, we have to work a little harder. Previous approaches involve  the dynamics of some probe that moves in the background of an instanton configuration.  The original work of \cite{wittenadhm} considered a heterotic string, moving in the background of an instantonic 5-brane. In the Type II context, one typically looks at a D0-brane moving in the background of a D4-D8 system, with the D4-brane absorbed into the worldvolume of the D8-brane where it appears as an instanton \cite{douglas2}. In both of these situations, the instanton configuration is fixed, and the ADHM data now appears as parameters of the theory rather than as dynamical degrees of freedom\footnote{There are other approaches to extracting the instanton gauge field from D-branes. The ADHM construction can be viewed as tachyon condensation \cite{sen} in a system of D4-branes and anti-D4-branes \cite{koji}. Alternatively, one can perturbatively reconstruct the large distance instanton solution by looking at open string vertex operators in the D$p$-D$p+4$ brane system \cite{lerda}.}. 

\para
The purpose of this paper is to provide a slightly different perspective on the ADHM construction of the gauge field. We return to the D0-D4 system, but now with the addition  of a fixed, heavy quark, represented by a Wilson line. The quark is electrically charged while the instanton is magnetically charged. As the instanton moves in the presence of the quark, it experiences a Lorentz force law proportional to its gauge profile $A_i$.  We will show that, from the perspective of the D0-brane quantum mechanics, the computation of this force reproduces the ADHM construction of the gauge field. 

\para
Our approach also yields a generalisation of the ADHM construction. A more precise version of the above statement is that when the {\it centre of mass} of the instanton moves, the Wilson line exerts a Lorentz force proportional to the gauge profile. But there are many other ways in which an instanton configuration can move: the sizes or orientations or relative separations of instantons can change. In all of these cases, the fixed quark exerts a force on the instantons. We explain how to compute this force from the gauge theory, and then show that the D0-brane quantum mechanics provides a simple expression for this force in terms of the ADHM data. 

\para
More generally, this paper fits into a growing literature on understanding the dynamics of solitons in the presence of electric or magnetic impurities. Other recent work in this area includes \cite{cherkis,cherkis2,vimp,mimp,moore}.

 \section{Instantons and Wilson Lines}\label{isec}

We start by describing the physics from a purely field theoretic perspective. We will derive the Lorentz-like force experienced by an instanton in the presence of a Wilson line. In Section \ref{dsec} we will re-examine this from the viewpoint of D-branes and see how it is related to the ADHM construction. 

\para
We work with an $SU(N)$ Yang-Mills theory in $d=4+1$ dimensions. The action for the  gauge field $A_\mu$ and a single adjoint-valued scalar $\varphi$ is given by
\be
S_{YM} = \frac{1}{e^2} \int d^5 x \ {\rm Tr}\left( - \frac{1}{2} F_{\mu\nu} F^{\mu\nu} - {\cal D}_\mu \varphi {\cal D}^\mu \varphi \right)
\label{5dact}\ee
%
%Here $F_{\mu\nu}  =\partial_\mu A_\nu - \partial_\nu A_\mu -i[A_\mu,A_\nu]$ and ${\cal D}_\mu \sigma= \partial_\mu \sigma-i[A_\mu,\sigma]$.
This can be viewed as part of an action with either ${\cal N}=1$ or ${\cal N}=2$ supersymmetry (i.e. either eight or sixteen supercharges). The theory with ${\cal N}=2$ supersymmetry has further scalar fields which will not play a role in our story.

\para
If we set $\varphi=0$, this action admits static soliton solutions obeying the self-dual Yang-Mills equations
\be 
F_{ij} = \frac{1}{2} \epsilon_{ijkl}F_{kl} \label{instantonselfdual}
\ee
where $i,j=1,2,3,4$ run over spatial indices.  Solutions to these equations are instantons. (For  reviews, see \cite{inscalc,tasi,nick}). 
Instantons are classified by the winding number $k\in {\bf Z}^+$ of the gauge field on the 3-sphere at infinity and the mass of such an instanton solution is given by 
\be M_{\rm inst} = \frac{8\pi k}{e^2}\nn\ee
The general solution to the instanton equations has $4kN$ parameters. For well-separated instantons, these correspond to 4 positions, a scale size and $4N-5$ orientation modes within the $SU(N)$ gauge group for each instanton. We write the general solution as $A_i(x;X^\alpha)$ with $X^\alpha$, $\alpha=1,\ldots 4kN$ the coordinates on the instanton moduli space ${\cal M}_{k,N}$ which takes the form
\be {\cal M}_{k,N} \cong {\bf R}^4 \times \tilde{\cal M}_{k,N}\label{mkn}\ee
Here the ${\bf R}^4$ factor captures the centre of mass of the instanton configuration, while the $\tilde{M}_{k,N}$ factor captures the relative positions, scale sizes and gauge orientations of the instantons.

\subsubsection*{Dynamics of Instantons}

The dynamics of slowly moving  instantons in $4+1$-dimensions can be described using the moduli space approximation \cite{manton}. 
This means that we promote the collective coordinates $X^\alpha$ to become time-dependent variables  $X^\alpha(t)$ and restrict attention to these degrees of freedom. 

\para
However, there is a subtlety: as the instantons move, they generate a non-Abelian electric field $E_i = F_{0i}$ and this should obey Gauss' law, ${\cal D}_iE_i=0$. Typically this doesn't happen automatically. Instead, we must turn on $A_0$ to ensure that the Gauss' law constraint holds. To achieve this, we start by 
introducing a {\it zero mode} associated to each of the collective coordinates. This is defined as the derivative of the gauge field together with an accompanying gauge transformation,
\be \delta_\alpha A_i = \frac{\partial A_i}{\partial X^\alpha} - {\cal D}_i\Omega_\alpha\label{omega}\ee
By construction, the zero mode is a solution to the linearised self-dual Yang-Mills equation.  We require the compensating gauge transformations $\Omega_\alpha(x,X)$  to solve the background gauge fixing condition,
\be {\cal D}_i \,(\delta_\alpha A_i)=0\label{ibgauge}\ee
where the covariant derivative is evaluated in the instanton background. 
The utility of this choice comes when we look at the non-Abelian electric field. This is given by
\be E_i = \frac{\partial A_i}{\partial X^\alpha}\dot{X}^\alpha - {\cal D}_iA_0\nn\ee
As we mentioned above, the electric field must satisfy Gauss' law ${\cal D}_i E_i =0$. If we set
\be A_0 = \Omega_\alpha(x;X) \dot{X}^\alpha\label{aois}\ee
then we have $E_i = \delta_\alpha A_i\,\dot{X}^\alpha$ and Gauss' law is obeyed by virtue of the gauge fixing condition \eqn{ibgauge}.

\para
Substituting this ansatz for the electric field into the action \eqn{5dact} gives us a description of the dynamics in terms of a sigma-model on the instanton moduli space ${\cal M}_{k,N}$,
\be S_{\rm instanton} = \int dt\ \frac{1}{2}\,g_{\alpha\beta}(X)\,\dot{X}^\alpha\dot{X}^\beta\label{insact}\ee
where the metric on ${\cal M}_{k,N}$ is given by the overlap of zero modes
\be
g_{\alpha \beta} (X) = \frac{2}{e^2} \int d^4 x \ {\rm Tr}\, \left( \delta_\alpha A_i \, \delta_\beta A_i \right)
\label{instg}\ee
This metric has a number of special properties: it is hyperK\"ahler and inherits an $SO(4)\times SU(N)$ isometry from spatial rotations and gauge action of the underlying field theory. However, it is not geodesically complete. The instanton moduli space has singularities where the instantons shrink to zero size; understanding the physics of these singularities presumably requires knowledge of the UV completion of our theory. The moduli space approximation is valid provided that $e^2\ll \rho$, where $\rho$ is the size of any given instanton. We assume that this condition holds in the following.

\para
For many applications, the metric \eqn{instg} is the most important geometric quantity associated to the instanton moduli space. Here, however, we will be more interested in the object $\Omega_\alpha$. This is an $SU(N)$ connection over ${\cal M}_{k,N}$.  To see this, suppose that we have a class of instanton solutions $A_i(x;X)$ presented in some gauge. We perform a gauge transformation, $A_i'=g A_ig^{-1} + i g \partial_i g^{-1}$, where $g=g(x;X)$, which means that we allow for the possibility of performing different gauge transformations at different points of the moduli space. We now need to find a different compensating gauge transformation $\Omega'$ such that the gauge fixing condition \eqn{ibgauge} holds for our new solution. It is not hard to show that this new compensating transformation is given by
\be \Omega'_\alpha = g\Omega_\alpha g^{-1} + ig\partial_\alpha g^{-1}\nn\ee
which means that $\Omega$ can indeed be viewed as an $SU(N)$ gauge connection over ${\cal M}_{k,N}$ as claimed. 

\para
The connection over the ${\bf R}^4$ factor of the moduli space ${\cal M}_{k,N}$ in \eqn{mkn}, describing the centre of mass of the instanton,  is particularly straightforward. The gauge condition \eqn{ibgauge} is satisfied if we take
\be \Omega_i= - A_i(X)\label{omegaisa}\ee
In other words, the auxiliary gauge connection $\Omega$ coincides with the physical gauge connection $A_i$ over ${\bf R}^4$. With this choice, the translational zero mode is given by $\delta_j A_i = F_{ij}$ and Gauss' law is satisfied.

\para
The connection over the reduced moduli space $\tilde{\cal M}_{k,N}$ is generally non-trivial. Explicit formulae on the $k=1$, $N=2$ case can be found in \cite{tasi}.

\subsubsection*{Instantons and Wilson Lines}

We now add a heavy, stationary quark to our theory, sitting at the origin of space. This is usually achieved by the insertion of a Wilson line in the path integral of the form,
\be W_R = {\rm Tr}_R\,{\cal T}\exp\left( i\int dt\ (A_0(t)-\varphi(t))\right)\label{swilson}\ee
Here ${\cal T}$ stands for time ordering and $R$ specifies the $SU(N)$ representation of the quark. Both the gauge field and scalar are evaluated at the origin of space: $A_0(t) = A_0(\vec{x}=0,t)$ and $\varphi(t)=\varphi(\vec{x}=0,t)$. The fact that the gauge field is accompanied in the Wilson line by a scalar field is familiar in supersymmetric theories  \cite{malda,sjrey}; this ensures that the Wilson line is BPS, preserving half of the supercharges.

\para
For our purposes, it will prove useful to work with a slightly different representation of the Wilson line, one which is at heart more semi-classical. To this end, we introduce a single, complex quantum mechanical degree of freedom, $\chi(t)$, which sits at the origin of space and transforms in the fundamental representation of the gauge group. It couples to the gauge field and scalar field through the action,
\be S_\chi = \int dt\ {\chi}^\dagger (i\partial_t - A_0(t) + \varphi(t) + M)\chi\label{chiact}\ee
where $M$ is the energy scale needed to excite $\chi$ and should be taken to be large.

\para
If $\chi$ is placed in its ground state, then it plays no role at energies $E\ll M$.  Things become more interesting if we excite some number of the $\chi$ degrees of freedom.  In this case, there is a close relationship between the action \eqn{chiact} and the Wilson line \eqn{swilson}. (See, for example, the textbook \cite{wittenbook}). The idea is to first do the path integral over the $\chi$ fields, in a fixed background $A_0(x)$ and $\varphi(x)$, and only subsequently perform the path integral over the $d=4+1$ dimensional super Yang-Mills fields. The number of excitations of $\chi$ is specified by including $p$ insertions in the path integral,
\be  Z_p[A_0,\phi] = \frac{1}{p!}\int {\cal D}\chi^\dagger{\cal D}{\chi}\ \chi_{a_1}(+\infty)\ldots \chi_{a_p}(+\infty)\,{\chi}^\dagger_{a_1}(-\infty)\ldots {\chi}^\dagger_{a_p}(-\infty)\,e^{iS_{\chi}}\ \ \ \ \label{path}\ee
It is straightforward to evaluate this path integral directly. One finds
\be  Z_p[A_0,\varphi] =  W_R[A_0,\varphi]\label{wline}\ee
where the representation $R$ depends on whether $\chi$ are quantised as fermions or bosons. If $\chi$ is quantised as a fermion then $R$ is the  $p^{\rm th}$ anti-symmetric representation of $SU(N)$; if $\chi$ is quantised as a boson then $R$ is the $p^{\rm th}$ symmetric representation. A recent, detailed derivation of \eqn{wline} can be found, for example, in \cite{gomis1}.

\para
We would like to understand how instantons move in the presence of the Wilson line \eqn{swilson} or, equivalently, in the presence of the $\chi$ degrees of freedom. This problem is identical to the one solved in \cite{mimp} where the dynamics of monopoles in the presence of Wilson lines was derived. (A very similar problem also arose in \cite{vimp}, which discussed the dynamics of Abelian vortices in the presence of charged impurities).

\para
Using the representation of the Wilson line in terms of $\chi$ fields allows us to work classically: we need to solve the equations of motion arising from the action $S=S_{\rm YM} + S_\chi$ given by \eqn{5dact} and \eqn{chiact}. The $\chi$ fields source both $A_0$ and $\varphi$. 
However, if we set $A_0=\varphi$, then the equations of motion for the spatial gauge fields are unchanged. This means that the static instanton configurations obeying $F_{ij} ={}^\star F_{ij}$ remain solutions to the equations of motion in the presence of the coupling to $\chi$. Meanwhile, $\varphi$, and hence $A_0$, is determined by the equation of motion
\be   {\cal D}^2 \varphi = e^2 {\chi}\chi^\dagger \delta^4(x)\label{chiphi}\ee
where the covariant Laplacian ${\cal D}^2$ is  evaluated on the background of the instanton. This equation is reminiscent of the discussion of dyonic instantons in \cite{dyin}; both configurations describe the BPS superposition of instanton and electric charges.

\para
Because the static instanton solutions are unchanged by the presence of a Wilson line, the moduli space of instantons is again given by ${\cal M}_{k,N}$. However, the dynamics of the instantons is now described by a quantum mechanics involving the  collective coordinates $X^\alpha(t)$ coupled to impurities $\chi(t)$.  When the instantons move, the ansatz for the temporal gauge field \eqn{aois} must be replaced by
\be A_0= \Omega_\alpha\dot{X}^\alpha + \varphi\nn\ee
with $\varphi$ determined by \eqn{chiphi}. 
This satisfies Gauss' law ${\cal D}_i E_i$ to leading order in $e^2/\rho$ (which is sufficient for the moduli space approximation).  Substituting this new ansatz into the Yang-Mills action yields the action,
\be S _{\rm instanton}= \int dt\ \frac{1}{2}\,g_{\alpha\beta}(X)\,\dot{X}^\alpha\dot{X}^\beta  +  \chi^\dagger(i\partial_t  + \Omega_\alpha(X)  \dot{X}^\alpha )\chi\label{ianswer} \ee
where the metric $g_{\alpha\beta}(X)$ is the same as that defined in \eqn{instg} and the connection is given by $\Omega(X) = \Omega(\vec{x}=0;X)$.  This second term gives the promised Lorentz force which couples the moving instanton to the Wilson line. If the instanton configuration moves rigidly, changing only its centre of mass then, by \eqn{omegaisa}, the force is governed by the Yang-Mills profile $A_i(x=0;X)$ as expected. (Note that knowledge of $A_i$ at $x=0$ is sufficient to reconstruct $A_i$ at all values of $x$; this is because if we decompose the moduli into the centre of mass coordinates,  $X$, and the remainder $\tilde{X}$, then gauge field has dependence $A_i(x-X;\tilde{X})$). 
However, this action also captures the dynamics if the instantons are undergoing a more complicated motion, changing their orientation, size or relative separation.

\para
The Wilson line and the instantons are half-BPS, and correspondingly, the quantum mechanics \eqn{ianswer} admits a supersymmetric completion with $\mathcal N = (0,4)$ supersymmetry. The constraints of supersymmetry provide a particularly simple derivation of some of the properties of the geometric quantities. They tell us, for example, that the metric $g_{\alpha\beta}$ must be hyperK\"ahler (which it is). They also tell us that the curvature for the connection $\Omega_\alpha$ is a $(1,1)$-form with respect to each of the three complex structures of the hyperK\"ahler manifold.

\section{D-branes and Wilson Lines}\label{dsec}

In this section, we provide a D-brane representation of instantons interacting with Wilson lines using the familiar D0-D4 system. Starting from the $U(k)$ quantum mechanics on the D0-branes, we will derive a low-energy effective action of the form \eqn{ianswer}. As we will see, the $SU(N)$ gauge connection $\Omega_\alpha$ --- which includes, as a special case, the Yang-Mills field $A_i$ --- will arise through the ADHM construction of the gauge connection. 

\para
Take $N$ D4-branes lying in the $x^{0,1,2,3,4}$ directions of Type IIA string theory. The worldvolume theory is described by $d=4+1$ dimensional, $U(N)$ Yang-Mills theory with ${\cal N}=2$ supersymmetry. These branes sit at the origin of ${\bf R}^5$, spanned by $x^{5,6,7,8,9}$.

\para
\EPSFIGURE{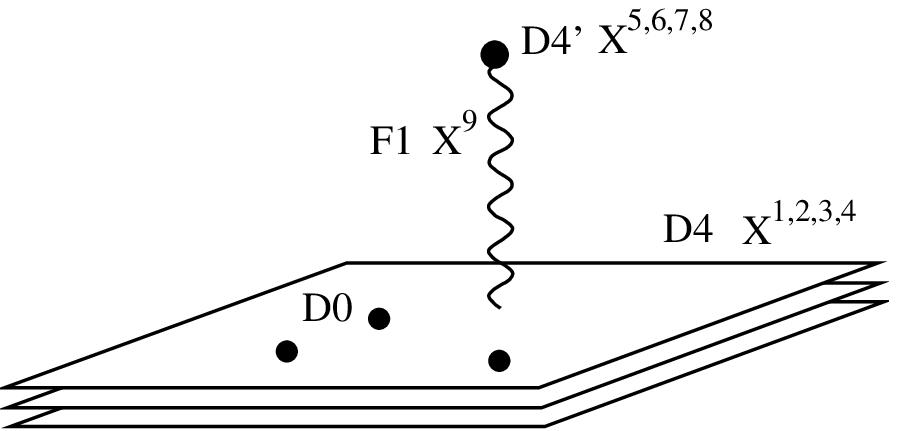,height=90pt}{}
We wish to add to this a Wilson line. This can be done by inserting an infinitely long fundamental string, lying in the $x^9$ direction as shown in the figure.  However, there is a simple trick, first introduced in \cite{gomis1},  that we will find useful. Rather than consider an infinitely long string, we will allow it to terminate on another D4-brane which we will denote as a D4$'$-brane.  Since we want the end of the string to be fixed, the D4$'$-brane should have worldvolume directions $x^{0,5,6,7,8}$. We place the D4$'$-brane at a distance $L$ in the $x^9$ direction. Ultimately we will take $L$ large.

\para
The string stretched between the D4-branes and the D4$'$-brane has 8 Dirichlet-Neumann directions. Upon quantisation, the lowest lying modes consist only of four complex fermions that we will call $\chi$. These lie in an ${\cal N}=(0,8)$ Fermi multiplet as described, for example, in \cite{evanati}. Each of these fermions transforms in the fundamental representation  of the $U(N)$ gauge group, with their dynamics governed by the action \eqn{chiact}, where the mass is given by $M= L/\alpha'$. There are also  couplings to the fields on the D4$'$-brane which we suppress since they won't be important in what follows\footnote{If the D4$'$-brane is wrapped on a compact space, this coupling becomes important since Gauss' law restricts the possible excitations of the $\chi$ fields. The introduction of a Chern-Simons coupling on the D4$'$-brane of the form $S_{\rm CS} = p \int dt\ A_0'$ with $p\in \mathbb{Z}$ ensures that the path integral is non-vanishing only when accompanied by $p$ field insertions as in \eqn{path}. For the non-compact D4$'$-brane considered here, the flux can escape to infinity and no such restriction applies.}. 

\para
As we reviewed in Section \ref{isec}, integrating out the fermions $\chi$ with $p$ excitations in the path integral is equivalent to the insertion of a Wilson line transforming in the $p^{\rm th}$ anti-symmetric representation. In the present context of D-branes, this observation was first made in \cite{gomis1}.

\subsection*{D0-Branes and Wilson Lines}

We now add $k$ D0-branes to our set-up. These appear as instantons when absorbed in the D4-branes. Our goal  is to understand how these D0-branes interact with the $\chi$ degrees of freedom arising from the D4-D4$'$ strings. 

\para
In fact, a T-dual version of this problem was solved recently in \cite{ads3}, where the $d=1+1$, ${\cal N}=(0,4)$ gauge theory describing  the D1-D5-D5$'$ system was constructed. We need only dimensionally reduce this theory to $d=0+1$ dimensions. The result is an ${\cal N}=(0,4)$ $U(k)$ gauged quantum mechanics. The theory has a $U(N)$ flavour symmetry and a $G = SO(4)^-\times SO(4)^+$ symmetry arising from rotations of the spatial ${\bf R}^4$ worldvolumes of the  D4 and D4$'$-branes respectively.  We write this as 
\be  G = SU(2)_L^-\times SU(2)_R^-\times SU(2)_L^+\times SU(2)_R^+\nn\ee
The two $SU(2)_R$ factors are R-symmetries of the superalgebra. 

\para
The field content of the theory is
\begin{itemize} 
\item D0-D0 strings: The D0-branes alone give rise to the familiar 16-supercharge $U(k)$ quantum mechanics, with all fields in the adjoint representation of the gauge group. There is a gauge field, $u_0$, and 9 real scalar fields which naturally decompose into two groups of 4 with one left over. The positions of the D0-branes in $x^{1,2,3,4}$ are denoted as  $Z^i$, with $i=1,2,3,4$. They transform under $SO(4)^-$. The positions of the D0-branes in $x^{5,6,7,8}$ are denoted as  $Y^i$, with $i=1,2,3,4$. They transform under $SO(4)^+$. Finally, the real scalar $W$ describes the positions of the D0-branes in the $x^9$ direction. 
There are also 8 complex, adjoint fermions, $\lambda$,  transforming under $G$ as $({\bf 1},{\bf 2},{\bf 1},{\bf 2})\oplus ({\bf 1},{\bf 2},{\bf 2},{\bf 1})\oplus({\bf 2},{\bf 1},{\bf 1},{\bf 2})\oplus({\bf 2},{\bf 1},{\bf 2},{\bf 1})$.

\item D0-D4 strings: The interaction with the D4-branes gives rise to an ${\cal N}=(4,4)$ hypermultiplet,   transforming as $({\bf k},\bar{\bf N})$ under the $U(k)\times U(N)$ symmetry. There are two complex scalars which we write as the doublet $\omega^T = (\phi,\tilde{\phi}^\dagger)$ transforming as $({\bf 1},{\bf 2},{\bf 1},{\bf 1})$ under $G$. The four complex fermions, which we denote collectively as $\psi$, transform as $({\bf 1},{\bf 1},{\bf 1},{\bf 2})\oplus ({\bf 1},{\bf 1},{\bf 2},{\bf 1})$.

\item D0-D4$'$ strings: The interaction with the D4$'$-brane gives rise to the same field content as the hypermultiplet, but with the interactions  twisted such that $SO(4)^+$ and $SO(4)^-$ are exchanged. The fields transform in the $({\bf k},{\bf 1})$ of $U(k)\times U(N)$.
The doublet of complex scalars $\omega^{\prime\,T} = (\phi',\tilde{\phi}^{'\,\dagger})$ transform as $({\bf 1},{\bf 1},{\bf 1},{\bf 2})$ under $G$. The four complex fermions, which we denote collectively as $\psi'$, transform as $({\bf 1},{\bf 2},{\bf 1},{\bf 1})\oplus ({\bf 2},{\bf 1},{\bf 1},{\bf 1})$. 

\item D4-D4$'$ strings: As we have seen, the D4-D4$'$ strings give rise to an ${\cal N}=(0,8)$ fermi multiplet, consisting of four fermions $\chi$. These transform as $({\bf 1},{\bf N})$ under $U(k)\times U(N)$ and are singlets under $G$.
\end{itemize}
The only difficultly in constructing the interactions of the theory is to determine how the $\chi$ fields couple to the rest. Perhaps surprisingly, it turns out that the coupling is uniquely fixed by supersymmetry: the interactions of the D0, D4 and D4$'$-branes alone do not preserve any supersymmetry\footnote{The field content is consistent with ${\cal N}=(0,4)$ supersymmetry but the interactions are not. The problem arises in ${\cal N}=(0,2)$ superfield language through the requirement that the two different kinds of superpotential obey a constraint ``$E\cdot J=0$". For more details, see \cite{ads3}.}. This can be rectified by an essentially unique (up to field redefinitions) interaction between the $\chi$ fermions and the D0-D4 and D0-D4$'$ strings. The full Lagrangian was given in \cite{ads3}; here we describe those couplings that are relevant for our story.

\subsubsection*{The Higgs Branch as the Instanton Moduli Space}

The scalar potential is most simply written by first introducing two triplets of D-terms, 
\be \vec{D}_Z= \vec{\eta}_{ij}Z^iZ^j + \omega^\dagger \vec{\sigma}\omega\ \ \ \ {\rm and}\ \ \ \ \  \vec{D}_Y = \vec{\eta}_{ij}Y^iY^j + \omega^{\prime\,\dagger} \vec{\sigma}\omega'\label{idterms}\ee
Here $\vec{\eta}$ are the self-dual 't Hooft matrices and $\vec{\sigma}$ are the Pauli matrices. Each of these looks like the triplet of D-terms that usually arises in theories with eight supercharges. The only novelty is that we now have a pair of these D-terms. The scalar potential is  given by
\be V &=& {\rm Tr}\,(\vec{D}_Z\cdot \vec{D}_Z + \vec{D}_Y\cdot \vec{D}_Y) + {\rm Tr}[Z^i,Y^j]^2 + \omega^\dagger (Y^iY^i +WW)\omega  \nn\\ && \ \ \ \  +\omega^{\prime\,\dagger}(Z^iZ^i-(W-M)(W-M))\omega'  + {\rm Tr}\,\left(\omega^\dagger\cdot \omega\,\omega^{\prime\,\dagger}\cdot\omega^{\prime}\right)\nn\ee
The indices in the last of these terms are constructed so that the expression is a singlet under the $U(N)$ and  $G$ but transforms in the adjoint representation under the $U(k)$ gauge group.

\para
The phase of the theory in which the D0-branes appear as instantons is characterised by  the requirement that $W=Y^i=\phi'=\tilde{\phi}'=0$, while $Z^i$, $\phi$ and $\tilde{\phi}$ are constrained to obey the D-term constraint $\vec{D}_Z=0$ as given in \eqn{idterms}.
There are $4k^2+ 4kN$ degrees of freedom in $Z$, $\phi$ and $\tilde{\phi}$. The D-terms above give $3k^2$ constraints. After dividing out by $U(k)$ gauge transformations, we're left with a $4kN$-dimensional space. This is the Higgs branch of the gauge theory and is known to coincide with the instanton moduli space \eqn{mkn}.

\para 
 The Higgs branch inherits a metric from the scalar kinetic terms and part of the ADHM construction is  the statement that this agrees with the metric \eqn{instg} on the instanton moduli space. 
  Although the construction of the Higgs branch metric is well known, it involves an ingredient that we will need later and, for this reason, we review it here. We introduce coordinates  $X^\alpha$, $\alpha=1,\ldots,4kN$ on the Higgs branch. This means that we can think of solutions to $\vec{D}_Z=0$  as being of the form $\omega(X)$, $Z^i(X)$.  Now suppose that we move on the Higgs branch, so $X\rightarrow X(t)$. We need to satisfy the constraint of Gauss' law within the $U(k)$ gauge theory on the D0-branes. If all other fields are set to zero, this reads,
\be i[Z^i,{\cal D}_0 Z^i] + i\omega{\cal D}_0\omega^\dagger - i({\cal D}_0\omega)\omega^\dagger =0\nn\ee
To solve this, we need to turn on the worldline $U(k)$ gauge field $u_0$. The way we do this is entirely analogous to the construction of the metric on the instanton moduli space. We first associate a zero mode to each degree of freedom. We define
 \be \hat\delta_\alpha \omega  = \frac{\partial \omega}{\partial X^\alpha} - i v_\alpha \omega \ \ \ ,\ \ \ \hat \delta_\alpha Z^i = \frac{\partial Z^i}{\partial X^\alpha} - i [v_\alpha,Z^i]\nn\ee
where, in each case, the change of the field is accompanied by an infinitesimal $U(k)$ gauge transformation,  $v_\alpha$. This compensating gauge transformation is determined by requiring that the zero modes obey a background gauge condition,
\be i[Z^i, \hat\delta_\alpha Z^i] + i\omega\hat\delta_\alpha\omega^\dagger - i(\hat\delta_\alpha\omega)\omega^\dagger =0\label{giggs}\ee
Then we can obey Gauss' law by setting the worldline gauge field $u_0$ to 
\be u_0 = v_\alpha \dot{X}^\alpha\label{uvwalt}\ee
This means that ${\cal D}_0Z = \hat \delta_\alpha Z\,\dot{X}^\alpha$ and ${\cal D}_0\omega = \hat \delta_\alpha \omega\,\dot{X}^\alpha$. Then, restricted to the Higgs branch, the low-energy dynamics of the quantum mechanics becomes
\be S_{\rm Higgs} &=& \int dt\  {\rm Tr} \,({\cal D}_0Z^i{\cal D}_0Z^i +{\cal D}_0\omega  {\cal D}_0\omega^\dagger  )    = \int dt\  \frac{1}{2}g_{\alpha\beta}(X)\,\dot{X}^\alpha\dot{X}^\beta\label{higgsact}\ee
Famously, the metric $g_{\alpha\beta}$ arising here coincides with the metric \eqn{instg}.  In this way, the D0-brane quantum mechanics captures the dynamics of instantons. The action has a supersymmetric completion which involves the fermions $\lambda$ and $\psi$.

\para
The upshot of this is that the instanton moduli space actually comes equipped with both a $U(k)$ connection $v_\alpha$ and an $SU(N)$ connection $\Omega_\alpha$. The D-brane picture naturally gives us the $v_\alpha$ connection. The field theory picture naturally gives us the $SU(N)$ connection. As we now show, the interaction with the Wilson line provides a natural map between them.

\subsubsection*{The Wilson Line}

We now turn to the dynamics of the fermions $\chi$ that are associated to the Wilson line. If we place all fermions $\chi$ in their ground state then they play no role in the dynamics. This corresponds to the situation where we do not excite any D4-D4$'$ strings. This situation is boring.

\para
Instead, we are interested in the case where the fundamental string between the stacks of D4-branes is in one of its lowest excited states. We pick one representative fermion, $\chi$, and choose to excite it just once. This  corresponds to a Wilson line in the fundamental representation of the $U(N)$ gauge group on the D4-branes. We would like to understand how this excited fermion affects the dynamics of the D0-branes.

\para
When we sit in the Higgs branch --- so that the D0-branes are absorbed in the D4-branes --- there is a mixing between $\chi$ and various other fermionic excitations. We can safely ignore any Yukawa couplings that involve $Y$, $W$, $\tilde{\phi}$ or $\tilde{\phi}'$ because these have been set to zero, but we're left with the following interactions 
\be L_{\rm Yukawa} = (\tilde{\psi}{}'_+,\bar{\psi}{}'_+)\left(\begin{array}{c}\phi \\ \tilde{\phi}{}^\dagger\end{array}\right)\chi + (\tilde{\psi}{}'_+,\bar{\psi}{}'_+) Z^i\sigma^i \left(\begin{array}{c} \bar{\tilde{\psi}}{}'_- \\ \psi'_-\end{array}\right)\label{yuk}\ee
Here the $\psi'$ fermions are those that arise from the D0-D4$'$ strings. We have distinguished four different kinds: $\psi'_\pm$ transform as ${\bf k}$ and $\tilde{\psi}'_\pm$ transform as $\bar{\bf k}$ under the $U(k)$ gauge group. The $\pm$ subscripts reflect the chirality of these fermions in $d=1+1$ dimensions before dimensional reduction. (The $\chi$ fermions are left-moving in $d=1+1$ and would be denoted $\chi_-$). 
The form of these Yukawa couplings is fully dictated by supersymmetry as described in \cite{ads3}.

\para
To understand the meaning of these Yukawa couplings, suppose for now that we place the two stacks of D4-branes on top of each other, so $M=0$ and there is no cost in energy to excite $\chi$. Then, on the Higgs branch, the Yukawa couplings above act as mass terms for the fermions. There are $2k$ ``right-moving" spinors and $2k + N$ ``left-moving" spinors. This means that, generically, we will be left with $N$ massless ``left-moving" spinors. (In the absence of any mixing due to the Yukawa terms, these are simply $\chi$). 

\para
Now separate the D4 and D4${}^\prime$ branes a distance $L$. It will cost energy $M=L/\alpha'$ to excite these $N$ modes, which are now a linear combination of $\chi$ and $\psi'$.  These excitations remain BPS excitations: they are the lowest energy excitations of the D4-D4${}^\prime$ strings and will  play the role of our Wilson line.  In order to determine which combination of $\chi$ and $\psi'$ we need to excite,  we introduce a $2k\times (N+2k)$ matrix, familiar to aficionados of the ADHM construction,
\be \Delta^\dagger = \left(\omega \, ,\, Z^i\bar{\sigma}^i\right)\nn\ee
where $\omega^T=(\phi,\tilde{\phi}{}^\dagger)$. This matrix is to be thought of as a function of the $4kN$ coordinates $X^\alpha$ which parameterise the Higgs branch.
Generically, this matrix has rank $2k$. We can introduce $N$ orthonormal, null eigenvectors, $U_a$, $a=1,\ldots,N$, each of them a $(N+2k)$-dimensional vector, defined by
\be \Delta^\dagger U_a = 0\ \ \ \ {\rm with}\ \ \ \ U_a^\dagger\cdot U_b =\delta_{ab}\label{udoesthis}\ee
Then the $N$ BPS modes, which receive no additional energy from the Yukawa interactions \eqn{yuk},  are given by 
\be \left( \chi  \, , \,\bar{\tilde{\psi}}{}^\prime_- \, ,\, \psi^\prime_-\right)^T = U_a\eta_a\nn\ee
where $\eta_a$ are $N$ Grassmann parameters. Restricted to these modes, the action for the three spinors becomes an action for the $\eta$ parameters, 
\be  S_\eta &=& i\bar{\chi}  \partial_0 {\chi} +i\bar{\tilde{\psi}}{}^\prime_-{\cal D}_0 {\tilde{\psi}}{}^\prime_-+i\bar{\psi}{}^\prime_-{\cal D}_0{\psi}{}^\prime_-
\nn\\ &=&i\bar{\eta}_bU^\dagger_b\left(U_a\partial_0{\eta}_a+({\cal D}_0 U_a )\eta_a\right) \nn\\ &=& \bar{\eta}_b\left(\delta_{ab}\,i\partial_0 +  (\Omega_\alpha)_{ab}\dot{X}^\alpha\right)\eta_a\label{etact}\ee
We see that the effective low-energy Lagrangian for the  Grassmann parameters $\eta$ contains an emergent  $U(N)$ gauge connection defined by
\be (\Omega_\alpha)_{ab} =iU_b{}^\dagger \hat \delta_\alpha U_a\label{adhmo}\ee
The covariant derivative $\delta_\alpha$ involves the $U(k)$ gauge connection $v_\alpha$ on the instanton moduli space. Its action on $U$ is a little unusual; tracing through the definitions above, we see that it only acts on the lower $2k$ components of the $(N+2k)$-vector $U$ which transform in the fundamental of $U(k)$.

 \para
 The computation of $\Omega$ above, which involved finding massless combinations of Yukawa couplings,  is very similar to that first introduced in \cite{wittenadhm}. However, there is an important difference: for us, the fields in $\Delta$ are dynamical rather than fixed parameters. This means that we end up with a connection over the full moduli space, ${\cal M}_{k,N}$, rather than just ${\bf R}^4$.

\para
The end result is an action for the D0-branes coupled to the Wilson line degrees of freedom,
\be
S=S_{\rm Higgs} + S_\eta\nn\ee
 with the two expressions given in \eqn{higgsact} and \eqn{etact}. This is to be compared with the gauge theory result \eqn{ianswer}. In both cases, the Wilson line degrees of freedom couple through an $SU(N)$ gauge connection that we have called $\Omega_\alpha$. It remains to show that these two $\Omega_\alpha$ are actually the same thing. This equivalence, as we now review, is the essence of the ADHM construction.

\subsection*{The ADHM Gauge Connection}

The  D-brane configuration naturally gives the $U(N)$ gauge connection  $\Omega$ over the instanton moduli space defined in \eqn{adhmo}. Let us first restrict attention to the ${\bf R}^4$ factor of ${\cal M}_{k,N}$ where, from the Yang-Mills analysis \eqn{omegaisa}, we expect $\Omega$ to coincide with the instanton gauge potential $A_i$. From the D0-brane quantum mechanics, it is simple to check that the associated $U(k)$ gauge connection vanishes in this case, $v_i=0$. Comparing the two, we  therefore expect that
\be
 A_i(X)= iU^\dagger \partial_i U\label{almostdone}\ee
Indeed, this is usually paraded as the key result of the ADHM construction \cite{adhm}. It is not hard to show that the field strength associated to $A_i$ is the self-dual instanton solution:  $F_{ij}={}^\star F_{ij}$. Simple proofs of this result can be found in any number of review articles such as \cite{inscalc,tasi,nick}.

\para
However, our D0-brane quantum mechanics has given us more.  When the instanton configuration moves in the relative moduli space $\tilde{\cal M}_{k,N}$, corresponding to changing the size, orientation or relative separations of the instantons, the force due to the Wilson line is captured by the other components of the  $U(N)$ connection  $\Omega_\alpha$. We would like  to show that this force is correctly captures the dynamics, meaning that the ADHM expression for $\Omega$ given in \eqn{adhmo} coincides with the definition introduced in Section \ref{isec}. In fact, this result  was first proven some time ago in  \cite{osborn} (and reviewed in \cite{inscalc}). For completeness, we describe the proof here.

\para
Let us recall what we want to show. The zero modes are defined by
\be \delta_\alpha A_i = \frac{\partial A_i}{\partial X^\alpha} - {\cal D}_i\Omega_\alpha\nn\ee
where $A_i$ is given in terms of ADHM data by \eqn{almostdone}. We claim that if  we take $\Omega_\alpha$ to be given by the expression \eqn{adhmo} then the zero mode automatically solves the background gauge fixing condition,
\be {\cal D}_i \,(\delta_\alpha A_i)=0\nn\ee
We start by writing down an expression for the zero mode in terms of ADHM data. In order to do this, we need the defining properties of $U$ \eqn{udoesthis} and $A_i$ \eqn{almostdone},  as well as the D-term constraints $\vec{D}_Z=0$ with $\vec{D}_Z$ given in  \eqn{idterms}. These D-term constraints have a nice consequence for $\Delta$  which can be shown to  satisfy the condition,
\be \Delta^\dagger \Delta = f^{-1} \otimes 1_{2\times 2}\nn\ee
for some invertible $k\times k$ matrix $f$. Making liberal use of these properties, one can show that the zero mode can be written as
\be\delta_\alpha A_i =U^\dagger {\cal M}_{\alpha i}U\nn\ee
where 
\be {\cal M}_{\alpha i} = i(\hat \delta_\alpha \Delta)\,f\,\partial_i\Delta^\dagger + {\rm h.c.}\nn\ee
Continuing down this road, one needs no new techniques and only a little more stamina, to find
\be
{\cal D}_i\, (\delta_\alpha A_j ) = U^\dagger \left(\partial_i {\cal M}_{\alpha j} - (\partial_i \Delta) f\Delta^\dagger {\cal M}_{\alpha j} - {\cal M}_{\alpha j} \Delta f (\partial_i \Delta^\dagger) \right) U\nn
\ee
Finally, we need to show that this vanishes when contracted over $i$ and $j$. This requires some obvious $\sigma$-matrix identities and, ultimately, reduces to the requirement that
\be {\rm Tr}_{2\times 2}\,(\Delta^\dagger (\hat \delta_\alpha \Delta) - (\hat \delta_\alpha \Delta^\dagger)\Delta)=0\nn\ee
%
%\be
% (f\otimes \sigma^i) (\Delta^\dagger D_\alpha \Delta) (f \otimes \bar \sigma^i) = 2 f\,  \rm{Tr}_{2 \times 2} (\Delta^\dagger D_\alpha \Delta) f \otimes 1_{2 \times 2}
%\ee
%
This is equivalent to our gauge fixing condition \eqn{giggs} for the Higgs branch zero modes. It is pleasing that the $U(k)$ gauge fixing on the Higgs branch implies the $U(N)$ gauge fixing condition. This concludes the proof that the ADHM connection $\Omega_\alpha$ defined in \eqn{adhmo} is the same object as the compensating gauge transformation defined in \eqn{omega}.

\section*{Acknowledgements}

We're grateful to Nick Dorey, Nick Manton, Andrew Singleton and David Skinner for very useful discussions and comments. We are supported by STFC grant ST/L000381/1 and by the European Research Council under the European Union's Seventh Framework Programme (FP7/2007-2013), ERC Grant agreement STG 279943, Strongly Coupled Systems. KW is supported by Gonville and Caius College.

\end{document}